\documentclass[aps,12pt,preprintnumbers,showpacs]{revtex4}

\usepackage{amsmath,amssymb}
\usepackage{bm}
\usepackage{comment}

\newcommand\Dlr{\raisebox{0.1em}{$\stackrel{\scriptstyle\leftrightarrow}D$}}
\renewcommand\d{\partial} \newcommand\x{\mathbf{x}}
 \newcommand\q{\mathbf{q}}
\newcommand\+{\dagger}

\begin{document}
\preprint{EFT-13-8}
\title{Newton-Cartan Geometry and the Quantum Hall Effect} 

\author{Dam Thanh Son}

\affiliation{Enrico Fermi Institute, James Franck Institute and
  Department of Physics, University of Chicago, Chicago, Illinois 60637,
  USA}

\begin{abstract}
We construct an effective field theory for quantum Hall states, guided
by the requirements of nonrelativistic general coordinate invariance
and regularity of the zero mass limit.  We propose Newton-Cartan
geometry as the most natural formalism to construct such a theory.
Universal predictions of the theory are discussed.

\end{abstract}
\maketitle

\section{Introduction}

The fractional quantum Hall (FQH)
states~\cite{Tsui:1982yy,Laughlin:1983fy} are the most nontrivial
states in condensed matter physics.  The observed quantization of the
Hall conductivity originates from nontrivial topological properties of
the ground state.  The topological nature of FQH states are also
expressed through the ground state degeneracy, nontrivial statistics
of quasiparticles, and edge modes.  What make the problem of the FQH
effect challenging is the crucial role played by interactions.

Many theoretical approaches have been suggested for the description of
the FQH states.  Two closely related approaches---the composite
boson~\cite{Zhang:1988wy} and composite fermion~\cite{Fradkin:1991wy}
approaches---have the advantages of being field theories, enabling
powerful theoretical tools.  However, one serious problem of these
approaches is the unnaturalness of the massless limit---the limit in
which the Coulomb energy scale is much smaller than the cyclotron
energy.  This problem exhibits itself in the tension between Kohn's
theorem and the existence of excitations at the Coulomb energy scales.

At energies much lower than the gap, it is usually believed that all
universal information about the quantum Hall states is described by a
pure Chern-Simons theory (``hydrodynamic theory'')~\cite{Wen:1992uk},
which encodes the quantized Hall conductivity.  There are, however,
universal properties related to transport beside the Hall
conductivity, at least in clean systems with rotation and Galilean
symmetries.  The Hall viscosity (also called odd viscosity or Lorentz
shear)~\cite{Avron:1995fg,TokatlyVignale:2007,Read:2009,TokatlyVignale:2008,ReadRezayi:2010}
is found to be a robust characteristic of gapped states and is
proportional to the shift~\cite{Wen:1992ej}. It has also been shown
that the $q^2$ correction to the Hall conductivity ($q$ being the
wavenumber of the longitudinal electric field) has a universal
coefficient which is related to the Hall
viscosity~\cite{Hoyos:2011ez,Bradlyn:2012ea}.  These universal
characteristics of the quantum Hall systems go beyond the framework of
the conventional hydrodynamic theory.

In this paper, we develop a field-theoretical formalism to capture
these new universal features of quantum Hall systems.  We will only
concern ourselves with physics at distance scales much larger than the
magnetic length, and energies far below the gap.  Universal results
derived in this paper are valid for the gapped FQH states with $\nu<1$
as well as the $\nu=1$ integer quantum Hall (IQH) state.  The
formalism makes use of a geometrical structure called the
Newton-Cartan geometry, which was originally developed for the purpose
of reformulating Newton's gravity in a coordinate invariant
way~\cite{Cartan1923,Kuenzle:1972zw,Ehlers}.  We found that the
mathematical machinery of the Newton-Cartan geometric structure is
particularly useful for developing an effective field theory
describing the quantum Hall states.

The structure of the paper is as follows.  In
Sec.~\ref{sec:symmetries} we review symmetry properties of the
microscopic theory, and the requirements for an effective theory
description.  Section~\ref{sec:NC} contains a brief overview of the
Newton-Cartan geometry.  In Sec.~\ref{sec:EFT} we construct the most
general effective theory consistent with the requirements put forward
in Sec.~\ref{sec:symmetries}.  The physical consequences are derived
in Sec.~\ref{sec:physics}.  Finally, Sec.~\ref{sec:conclusion}
contains concluding remarks.

\section{Symmetry considerations}
\label{sec:symmetries}

\subsection{Nonrelativistic diffeomorphism}

A system of nonrelativistic particles (electrons) has several
conservation laws.  It is well known that the conservation of particle
number is related to a gauge invariance of the action describing
electrons in an external electromagnetic field.  Similarly,
conservation of momentum is related to a gauge invariance in the
theory describing electrons interacting with the gauge field and the
metric.  Let us first consider the case of noninteracting particles.
We can couple the system to the external gauge field $A_\mu$ and the
metric $h_{ij}$ in the following way,
\begin{equation}\label{S}
  S
   =\!\int\!d^3x\, \sqrt h\, \left[ \frac i2
      \psi^\+ \Dlr_t\psi - \frac{h^{ij}}{2m}
      D_i \psi^\+ D_j\psi + \frac{gB}{4m} \psi^\+ \psi 
     \right],
\end{equation}
where $D_\mu=\d_\mu-iA_\mu$, $h^{ij}$ is the inverse of $h_{ij}$,
$h=\det h_{ij}$, $B = (\d_1 A_2 - \d_2 A_1)/\sqrt h$, and $g$ is the
g-factor of the electrons, assumed to be fully polarized.  The metric
is only a spatial metric; time is absolute.  At the end, we will be
mostly interested in physics occurring in flat space, but introducing
the metric turns out to be a useful intermediate step.  There are some
ambiguities in the coupling of matter to the metric, but for the
purpose of this paper, the simplest coupling~(\ref{S}) is sufficient.
The magnetic moment term in~(\ref{S}) modifies the dynamics only when
the magnetic field is inhomogeneous, but even when $B$ is constant it
modifies the expression for the current $j^i=\delta S/\delta A_i$,
adding to it a magnetization current.

The action is invariant under gauge transformations: $\delta \psi =
i\alpha\psi$, $\delta A_\mu = \d_\mu\alpha$.  By direct
computation, one can also verify that it is invariant with respect to
time-dependent general coordinate transformations, characterized by
the gauge parameters $\xi^k(t,\x)$,
\begin{subequations}\label{transf}
\begin{align}
  \delta\psi &= -\xi^k \d_k \psi, \\
  \delta A_0 &= -\xi^k \d_k A_0 -A_ k\dot \xi^k + 
   \frac g4 \varepsilon^{ij} \d_i(h_{jk}\dot \xi^k),
    \label{trans-A0}\\
  \delta A_i &= -\xi^k \d_k A_i -A_k\d_i \xi^k 
     - m h_{ik}\dot \xi^k, \label{trans-Ai}\\
  \delta h_{ij} & = -\xi^k \d_k h_{ij} - h_{kj}\d_i \xi^k 
    - h_{ik}\d_j\xi^k. \label{h-trans}
\end{align}
\end{subequations}
Here $\varepsilon^{ij}=\epsilon^{ij}/\sqrt h$, and $\epsilon^{ij}$ is
the totally antisymmetric symbol with
$\epsilon^{12}=-\epsilon^{21}=1$.  Equations~(\ref{transf}) correspond
to time-dependent coordinate transformations $x^k\to x^k+\xi^k(t,x)$.
The $g=0$ version of this invariance was considered previously in
Refs.~\cite{Son:2005rv,Hoyos:2011ez}.  In this case, the invariance
can be thought of as a nonrelativistic limit of a relativistic
coordinate invariance~\cite{Son:2005rv}.  It has also been shown that
interactions can be introduced to the system in a way which respects
the general coordinate invariance~\cite{Hoyos:2011ez}.

\subsection{Requirements for the effective theory}
\label{sec:reqs}

The problem of finding the electromagnetic and gravitational response
of a quantum Hall fluid is that of finding the effective action
$S[A_0, A_i, h_{ij}]$.  By the ``effective action'' here we simply
mean the generating functional that one would obtain, in the
path-integral formalism, if one was able to perform the path integral
over the electron field $\psi$.  However such direct integration is
feasible only for IQH states but not for FQH states.  We hence will
have to rely on general principles.  

Because the quantum Hall states are gapped, $S$ can be expanded in
Taylor series over powers of fields and derivatives.  Our goal is only
to find the lowest terms in the derivative expansion of $S$.

The first requirement is that $S$ is gauge invariant and general
coordinate invariant,
\begin{align}
  S[A_0+\d_0\alpha,\, A_i+\d_i\alpha,\, h_{ij}] &= S[A_0,\,A_i,\,h_{ij}],\\
   S[A_0 + \delta A_0,\, A_i + \delta A_i,\, h_{ij} + \delta h_{ij}] 
  &= S[A_0,\, A_i,\, h_{ij}] + O(\xi^2).
\end{align}
where in the second equation $\delta A_0$, $\delta A_i$, and $\delta
h_{ij}$ are given in Eqs.~(\ref{transf}).  Next, we note that $A_0$
enters the action~(\ref{S}) only through the combination $A_0+gB/4m$,
hence the electromagnetic responses of systems with different
g-factors are related.  Suppose $S_g[h_{ij}, A_0, A_i]$ is the
effective action determining the response of a system with g-factor
$g$.  Then
\begin{equation}\label{SgS2}
  S_g [A_0,\, A_i,\, h_{ij}] =
  S_{g'} \left[A_0 + \frac{g-g'}{4m}B, A_i,\, h_{ij}\right].
\end{equation}
Hence if one could find $S$ for one particular value of $g$, then one
would know $S$ for all $g$'s.

The special value of the g-factor is $g=2$.  At this value, the lowest
Landau level is completely degenerate at zero energy, even when the
magnetic field $B$ is inhomogeneous and the metric $h_{ij}$ is
nontrivial~\cite{Aharonov:1978gb,Maraner:1992sz,Alicki:1993nt}.  In
this case, if one sends all higher Landau levels to infinite energy by
taking $m\to0$, effective action describing states at the lowest
Landau level should remain finite.  Thus, another requirement is the
existence of a regular limit
\begin{equation}
  \lim_{m\to 0} S_{g=2}[h_{ij}, A_0, A_i].
\end{equation}
In particular, for any $g\neq2$ the limit $m\to0$ in $S$ is singular
(unless $S$ does not depend on $A_0$, which is unphysical).  We note
that the transformation laws~(\ref{transf}) are not singular in the
limit $m\to0$.

\subsection{The necessity to improve the standard hydrodynamic theory}

After integrating out all dynamical fields, the standard hydrodynamic
theory~\cite{Wen:1992uk} gives a Chern-Simons action involving the
external gauge potential,
\begin{equation}\label{SCS}
  S_{\rm CS}[A] = \frac\nu{4\pi}\int\!d^3x\, 
  \epsilon^{\mu\nu\lambda}A_\mu \d_\nu A_\lambda\, .
\end{equation}
This action encodes the Hall effect with Hall conductivity
$\sigma_{xy}=\nu/2\pi$ (in units of $e^2/\hbar$).  On the other hand,
we expect the effective theory to respect the symmetry~(\ref{transf})
of the microscopic theory.  Under the general coordinate
transformations~(\ref{trans-A0}), (\ref{trans-Ai}), the Chern-Simons
action changes,
\begin{equation}
  \delta S_{\rm CS}[A] = \frac\nu{2\pi}\!\int\!d^3x\, \varepsilon^{ij}
 \left(mE_i - \frac g4\d_iB\right) h_{jk}\dot\xi^k .
\end{equation}
So the action is not invariant under general coordinate
transformations unless we take the limit $m=0$ and the g-factor is
zero, $g=0$.  The root of the problem is that, except for this
particular case, $A_0$ and $A_i$ do not transform like the components
of a one-form.

Thus we conclude, from symmetries alone, that the Chern-Simons
action~(\ref{SCS}) cannot be the complete effective action for the
quantum Hall states for generic $m$ and $g$.  Can this action be the
complete action in the regime $g=0$, $m\to0$?  It is easy to argue
that it cannot be.  Indeed, as noted above, the effective action must
be singular in the limit $m\to0$ if $g\neq2$; at the same time,
(\ref{SCS}) is completely regular. Another way to say the same thing
is that, if (\ref{SCS}) was the effective action for $g=0$, the action
for $g=2$ would be, according to Eq.~(\ref{SgS2}),
\begin{equation}
  S_{g=2}[A] = \frac\nu{4\pi}\!\int\!d^3x\,\epsilon^{\mu\nu\lambda}
   A_\mu \d_\nu A_\lambda + \frac\nu{4\pi m}\!\int\!d^3x\,
   \sqrt h\, B^2 ,
\end{equation}
which becomes singular when $m\to0$, in contradiction with the
regularity of the $m\to0$ limit at $g=2$. Thus, (\ref{SCS}) cannot be
the complete action for any value of $g$ and $m$.

This conclusion may appear trivial, because generically one expects
that, after integration over $\psi$, terms with all numbers of
derivatives are generated.  However, by showing that (\ref{SCS}) does
not satisfy the general requirements, we anticipate that some of these
higher-derivative terms are completely fixed by symmetries and
regularity in the massless limit.

There is a second deficiency of the action~(\ref{SCS}): it does not
encode the shift and the Hall viscosity.  Before discussing a new,
improved action, we need to discuss the Newton-Cartan geometry that
underlies its construction.

\section{Newton-Cartan geometry}
\label{sec:NC}

The Newton-Cartan geometric structure first appeared in Cartan's
reformulation of Newton's gravity in coordinate-invariant
language~\cite{Cartan1923} and was subsequently developed by others
(see, e.g., Refs.~\cite{Kuenzle:1972zw,Ehlers}).  We give here a
short, self-contained summary of aspects of Newton-Cartan's gravity
relevant for this work, with special emphasis on the case of (2+1)
dimensions.

\subsection{The geometric structure}

A Newton-Cartan geometry is a
structure consisting of
\begin{itemize}
\item a manifold, on which one can choose any system of coordinates
  $x^\mu$, and where tensors are defined by transformation properties
  under coordinate change;
\item a degenerate metric $h^{\mu\nu}$ with one zero eigenvalue, and all
other eigenvalues being positive;
\item a one-form $n=n_\mu dx^\mu$, which is a closed form in the 
  torsionless version of Newton-Cartan geometry;
\item A vector $v^\mu$, called the velocity vector, which satisfies 
$n\cdot v=1$.
\end{itemize}
From $(h^{\mu\nu},n_\mu,v^\mu)$ one can define a unique metric tensor
with lower indices $h_{\mu\nu}$ by requiring
\begin{equation}
  h^{\mu\lambda} h_{\lambda\nu} = \delta^\mu_\nu -v^\mu n_\nu, \quad
  h_{\mu\nu} v^\nu =0.
\end{equation}

A symmetric connection can be introduced,
\begin{equation}\label{Gamma}
  \Gamma^\lambda_{\mu\nu} = v^\lambda \d_\mu n_\nu + \frac12
  h^{\lambda\rho}\left( \d_\mu h_{\nu\rho} + \d_\nu h_{\mu\rho} -\d_\rho
    h_{\mu\nu}\right).
\end{equation}
It is easy to check that~(\ref{Gamma}) transforms as required for a
connection.  Covariant derivative defined with the
connection~(\ref{Gamma}) possesses many interesting properties:
\begin{align}
  & \nabla_{\!\lambda} h^{\mu\nu} = 0, \quad \nabla_{\!\lambda} n_\mu  = 0,
     \quad
  h_{\alpha[\mu}\nabla_{\!\nu]}v^\alpha  = 0, \label{nablagnv} \\
  & v^\lambda\nabla_\lambda h_{\mu\nu} = 0, \quad 
    v^\lambda\nabla_\lambda v^\mu=0, \quad
    h^{\mu\alpha}h^{\nu\beta} \nabla_\lambda h_{\alpha\beta} = 0,
  \quad h^{\alpha[\mu}\nabla_\alpha v^{\nu]} = 0.
\end{align}
In fact, the connection~(\ref{Gamma}) is uniquely determined if one
requires three conditions in Eqs.~(\ref{nablagnv}).

The Newton-Cartan structure arises naturally from dimensional
reduction along a light-cone direction.  Consider a space with one
extra dimension, parameterized by the coordinates $x^M=(x^-, x^\mu)$,
and with the metric
\begin{equation}
  ds^2 = G_{MN}dx^M dx^N = 2 n_\mu dx^- dx^\mu + h_{\mu\nu} dx^\mu dx^\nu
\end{equation}
The metric of this space is not degenerate and so can be inverted,
\begin{equation}
  G_{MN} = \begin{pmatrix} 0 & n_\nu \\ n_\mu & h_{\mu\nu} \end{pmatrix},
  \quad
  G^{MN} = \begin{pmatrix} 0 & v^\nu \\ v^\mu & h^{\mu\nu} \end{pmatrix}.
\end{equation}
The Christoffel symbols $\Gamma^L_{MN}$, when indices are restricted
to those different from $x^-$, coincide with~(\ref{Gamma}).

The Newton-Cartan formalism allows equations to be written in any
system of coordinates.  However, there is a special class of
coordinate systems where the time $x^0$ is chosen to be the ``global
time.''  The global time $t$ is defined through $n$: $n=dt$ (recall
that $n$ is a closed one-form).  We will call any coordinate system
where $t$ is chosen as the time coordinate, $x^0=t$, a global-time
coordinate system.  Note that after fixing $x^0=t$, there is still a
freedom of choosing the spatial coordinates $x^i$.  This gauge freedom
is parameterized by the functions $\xi^i(t,\x)$, corresponding to
$x^i\to x^{i\prime} = x^i+\xi^i(t,\x)$.

In global-time coordinate systems the components of $n_\mu$ are
$n_\mu=(1,\vec 0)$.  Due to $h^{\mu\nu}n_\nu=0$, in such a coordinate
system the components of $h^{\mu\nu}$ are
\begin{equation}\label{h-GT}
  h^{\mu\nu} = \begin{pmatrix} 0 & 0 \\ 0 & h^{ij} \end{pmatrix} .
\end{equation}
The velocity $v^\mu$ and $h_{\mu\nu}$ can be parameterized through the
spatial components of the velocity, $v^i$,
\begin{equation}
  v^\mu = \begin{pmatrix} 1 \\ v^i \end{pmatrix}, \quad
  h_{\mu\nu} = \begin{pmatrix} v^2 & -v_j \\ -v_i & h_{ij} \end{pmatrix},
\end{equation}
where, $h_{ij}$ is the inverse matrix of $h^{ij}$ and, for notational
convenience, we denote $v_i=h_{ij}v^j$ and $v^2=v^i v_i$. (Note: $v_i$
are not the spatial components of a spacetime co-vector, and $v^2$ is
not the square of a spacetime vector).

Under spatial reparameterization ($\xi^0=0$) the form~(\ref{h-GT}) is
preserved while $h_{ij}$ transforms as
\begin{equation}
  \delta h_{ij} = -\xi^k \d_k h^{ij} - h_{ik}\d_j\xi^k -h_{kj}\d_i\xi^k .
\end{equation}
Notice that this is the same as Eq.~(\ref{h-trans}).  Later in our
construction of the Newton-Cartan geometry of the quantum Hall states,
the external metric $h_{ij}$ will play the role of the spatial metric
of the geometry.  The spatial components of the velocity vector
$v^i$ transform as
\begin{equation}\label{trans-vi}
  \delta v^i = - \xi^k \d_k v^i + v^k \d_k \xi^i + \dot \xi^i .
\end{equation}
The last term in Eq.~(\ref{trans-vi}) justifies calling $v^i$ the
``velocity.'' For example, under a Galilean boost with $\xi^i=V^i t$,
one has $v^i\to v^i+V^i$.

The Newton-Cartan geometric structure can be visualized as a
collection of Riemannian spaces, one space at each moment of time,
with a spatial metric on each time slice and with a velocity field
connecting Riemannian spaces at different times.  Parallel transport
within a time slice can be done with the use of the metric $h_{ij}$,
but parallel transport from one time slice to another requires the
velocity vector $v^i$.

\subsection{The shear tensor}

The shear tensor $\sigma_{\mu\nu}$ can be defined without using the
connection as
\begin{equation}
  \sigma_{\mu\nu} = \pounds_v h_{\mu\nu} = v^\lambda\d_\lambda h_{\mu\nu}
   + h_{\lambda\nu}\d_\mu v^\lambda + h_{\mu\lambda}\d_\nu v^\lambda .
\end{equation}
The shear tensor is symmetric and satisfies $v^\mu\sigma_{\mu\nu}=0$.
The covariant derivatives of $v^\mu$ and $h_{\mu\nu}$ then can be
expressed in terms of the shear tensor,
\begin{equation}
  \nabla_\mu v^\nu =  \frac12\sigma_{\mu\lambda} h^{\lambda\nu}, \quad
  \nabla_\lambda h_{\mu\nu} = - \sigma_{\lambda(\mu} n_{\nu)} .
\end{equation}
In a global-time coordinate system, the spatial components of the
shear tensor are
\begin{equation}
  \sigma_{ij} = \nabla_i v_j + \nabla_j v_i + \dot h_{ij},
\end{equation}
where the covariant derivatives in the last equation are defined with
respect to the spatial metric $h_{ij}$.  This justifies the name
``shear tensor.'' The other components of $\sigma_{\mu\nu}$ are
uniquely fixed by $v^\mu\sigma_{\mu\nu}=0$.  From $\sigma_{\mu\nu}$ we
can construct the traced and traceless part,
\begin{equation}
  \sigma = h^{\mu\nu}\sigma_{\mu\nu} = 2\nabla_\mu v^\mu, \quad
  \hat\sigma_{\mu\nu} = \sigma_{\mu\nu} - \frac1d h_{\mu\nu}\sigma,
\end{equation}
where $d$ is the number of spatial dimensions.  In the rest of this
paper we take $d=2$.

\subsection{The spin connection}

The spin connection plays an important role in our construction of the
effective action for the quantum Hall state.  We assume the
Newton-Cartan space is $(2+1)$ dimensional, Let us define at each
point a pair of vectors (a vielbein) $e^{a\mu}$, $a=1,2$ so that
$n_\mu e^{a\mu}=0$ and
\begin{equation}
   h^{\mu\nu} = \sum_{a=1}^2 e^{a\mu} e^{a\nu} .
\end{equation}
By lowering the index we can define $e^a_\mu= h_{\mu\nu}
e^{a\nu}$ with the properties
\begin{equation}
  h_{\mu\nu} = \sum_{a=1}^2 e^a_\mu e^a_\nu, \quad v^\mu e^a_\mu =0 .
\end{equation}
We will also chose to orient the basis vectors $e^a$ so that
$\epsilon^{\lambda\mu\nu} \epsilon^{ab} n_\lambda e^a_\mu e^b_\nu>0$.

The spin connection can be defined as 
\begin{equation}
  \omega_\mu = \frac12 \epsilon^{ab} e^{a\nu} \nabla_{\!\mu} e^b_\nu .
\end{equation}

In global-time coordinate systems, the vielbein vectors have
components
\begin{equation}
  e^{a\mu} = (0, \,  e^{ai}) , \quad
  e^a_\mu = (- v_i e^{ai}, \, e^a_i ),
\end{equation}
and the components of the spin connection are
\begin{align}
  \omega_0 &= \frac12\left(\epsilon^{ab} e^{aj}\d_0 e^b_j + 
    \varepsilon^{ij} \d_i v_j\right) , \\
  \omega_i &= \frac12\left(\epsilon^{ab} e^{aj}\d_i e^b_j 
    - \varepsilon^{jk} \d_j h_{ki}\right).
\end{align}
The last term in $\omega_0$ is the vorticity if $v_i$ is interpreted
as the velocity field of a flow.  The spin connection $\omega_\mu$
transforms like an abelian gauge field under O(2) local rotation of
the vielbein $e^a$.  The field strength tensor
$\omega_{\mu\nu}=\d_\mu\omega_\nu-\d_\nu\omega_\mu$ is independent of
the choice of the vielbein; its spatial component is the scalar
curvature:
\begin{equation}\label{dwR}
  \omega_{12}=\d_1\omega_2-\d_2\omega_1=\frac12\sqrt h R.
\end{equation}

\section{Effective field theory of the quantum Hall state.}
\label{sec:EFT}

\subsection{Improved gauge potentials}

Let us recall that~(\ref{SCS}) does not respect diffeomorphism
invariance, since $A_\mu$ does not transform as a one-form.  To write
down a diffeomorphism invariant action, we imagine the quantum Hall
state to live in a Newton-Cartan geometry.  The external metric
$h_{ij}$ transforms correctly under spatial coordinate transformations
and hence can be taken as the metric of the Newton-Cartan geometric
structure.  We lack, however, a ready-made velocity field $v^\mu$, but
for a moment let us assume that such a field has somehow emerged
dynamically.  With $v^\mu$ at hand, let us consider the following
object,
\begin{align}\label{u0ui}
  \tilde A_0 &= A_0 - \frac  m2 v^2 - \frac g4
    \varepsilon^{ij}\d_i v_j , \\
  \tilde A_i &= A_i + m v_i ,
\end{align}
where $v_i = h_{ij} v^j$, $v^2=v^i v_i$.  This object is interesting
for the following reason.  First, under gauge transformations, $\tilde
A_\mu$ transforms like a gauge potential: $\tilde A_\mu\to \tilde
A_\mu + \d_\mu\alpha$.  Second, using Eqs.~(\ref{transf}) and
(\ref{trans-vi}), we find that $\tilde A_\mu$ transforms like a
one-form under diffeomorphism,
\begin{equation}
  \delta \tilde A_\mu = - \xi^\lambda\d_\lambda \tilde A_\mu 
   - \tilde A_\lambda \d_\mu \xi^\lambda ,
\end{equation}
where $\xi^\mu=(0,\xi^i)$.  Thanks to these properties, $\tilde A_\mu$
will be especially useful for our future discussion.

\subsection{A Chern-Simons effective action}

We will search for an action $S[A_\mu, h_{ij}]$ with the required
symmetry properties.  First we perform a Legendre transform of the
action with respect to the transverse part of $A_\mu$ to write
\begin{equation}\label{SSj}
  S[A_\mu, h_{ij}] = S_j[j^\mu, h_{ij}] - \int\!d^3x\, \sqrt h\, 
  j^\mu (\d_\mu\varphi - A_\mu).
\end{equation}
$S_j[j^\mu, h_{ij}]$ contains the same amount of information as
$S[A_\mu, h_{ij}]$.  Extremizing the right hand side of Eq.~(\ref{SSj}) with
respect to $j^\mu$ and $\varphi$ we should obtain the functional 
$S[A_\mu,h_{ij}]$.
We note that, since the action $S[A_\mu,h_{ij}]$
contains a Chern-Simons term, the action $S_j$ contains a nonlocal
contribution.  To separate out this contribution we introduce a gauge
field $a_\mu$ and rewrite the action as
\begin{equation}\label{Sjloc}
  S = \frac\nu{4\pi}\!\int\!d^3x\,\epsilon^{\mu\nu\lambda}a_\mu\d_\nu a_\lambda
      - \int\!d^3x\, \sqrt h\, j^\mu (\d_\mu\varphi - A_\mu + a_\mu)
      + S_j^{\rm loc}[j^\mu, h_{ij}] ,
\end{equation}
where $S_j^{\rm loc}$ is now a local functional of its variables.
Next denote $j^\mu=\rho v^\mu$, $v^\mu=(1,v^i)$.  Since $j^\mu$
transforms like a vector, $v^\mu$ is also a vector.  The
action~(\ref{Sjloc}) is still not written in an explicitly invariant
form since $A_\mu$ does not transform like a one form.  Thus we
separate out a part from $S_j^{\rm loc}$,
\begin{equation}
  S_j^{\rm loc}[j^\mu, h_{ij}] = S'[\rho, v^i, h_{ij}] 
    + \int\!d^3x\, \sqrt h\,\rho \left( 
  \frac{mv^2}2 - \frac g4 \varepsilon^{ij}\d_i v_j\right).
\end{equation}
Then Eq.~(\ref{Sjloc}) becomes
\begin{equation}\label{SSpr}
  S = \frac\nu{4\pi}\!\int\!d^3x\,\epsilon^{\mu\nu\lambda}a_\mu\d_\nu a_\lambda
      - \int\!d^3x\, \sqrt h\, \rho v^\mu (\d_\mu\varphi - \tilde A_\mu
      + a_\mu) + S'[\rho, v^i, h_{ij}] .
\end{equation}
Since the two first terms are diffeomorphism invariant, $S'[\rho, v^i,
  h_{ij}]$ should also be diffeomorphism invariant.

\subsection{Coupling of composite boson to spin connection}

We now show that there is a universal contribution to $S'$ responsible
for the shift and the Hall viscosity of the quantum Hall liquid.
Before giving a general argument, it is instructive to go over a
heuristic argument based on the flux attachment procedure.

Assume we are dealing with a Laughlin fraction, where $\nu=1/(2p+1)$.
The variable $\varphi$ in Eq.~(\ref{SSpr}) can be interpreted as the
phase of the condensate of ``composite bosons,'' obtained from
attaching $1/\nu=2p+1$ flux quanta to the original fermions.  We now
argue that in a curved background, such composite bosons should couple
to the metric through the spin connection.

Let us recall that the shift $\mathcal S$ is defined as the offset in
the linear relationship between the number of particles in the ground
state of a quantum Hal state and the number of magnetic flux quanta
$N_\phi$: $N_\phi=\nu^{-1}N-\mathcal S$~\cite{Wen:1992ej}.  For
definiteness let us take $\nu=1/3$, where $\mathcal S=1/\nu=3$, and
consider the ground state on a sphere.  For $3N$ flux quanta through
the sphere, there are $N+1$ electrons in the ground state.  Let us
perform the standard flux attachment procedure, attaching $-3$
statistical flux quanta to each electron.  The total flux through the
sphere is now $3N-3(N+1)=-3$.  This seems to contradict the fact that
the composite bosons form a condensate without any vortices.  (Note
that this problem does not arise on a torus.) To resolve this problem,
one needs to assume that the composite boson is coupled to the spin
connection, so that the curvature of the sphere supplies the missing 3
flux quanta.  The total curvature flux through a sphere is 2, so the
composite boson should carry charge $3/2$ with respect to the spin
connection.

Thus we have found that for a general Laughlin's fraction
$\nu=1/(2p+1)$, the spin connection charge of the composite boson is
$s=1/2\nu$.  This means that the covariant derivative of the
condensate phase $\varphi$ should be defined as
\begin{equation}\label{Dphi}
  D_\mu \varphi = \d_\mu \varphi - \tilde A_\mu - s\omega_\mu + a_\mu.
\end{equation}

We now present the general argument.  If $\varphi$ couples to the
gauge field and the metric as in Eq.~(\ref{Dphi}), then on a closed
manifold, the field $\varphi$ can be free of singularities only if the
total flux of the gauge field coupled to it,
$A_\mu+s\omega_\mu-a_\mu$, through the whole space is zero:
\begin{equation}
  \int\!d^2x\, \sqrt g \left( B + \frac s2 R - b \right) = 0,
\end{equation}
where we have used Eq.~(\ref{dwR}).  Notice, however, that variation
of the action~(\ref{SSpr}) with respect to $a_0$ implies $\rho=\nu
b/2\pi$.  Thus, the equation above reads $N_\phi+s\chi-\nu^{-1}N=0$,
where $\chi$ is the Euler characteristics of the manifold.  Since
$\chi=2$ for a sphere, the relationship between $s$ and the shift
$\mathcal S$ is $s=\mathcal S/2$.

\subsection{The action}

From the previous discussion, we separate a term proportional to $\rho
v^\mu \omega_\mu$ from $S'$ and write the action in the final form,
\begin{equation}\label{Sg2}
  S =  \!\int\!d^3x\ \bigl( \frac\nu{4\pi} 
   \epsilon^{\mu\nu\lambda}a_\mu\d_\nu a_\lambda 
 - \sqrt h\, \rho v^\mu D_\mu\varphi \bigr) + S_0[\rho, v^i, h_{ij}],
\end{equation}
with $D_\mu\varphi$ defined in Eq.~(\ref{Dphi}).

So far we have not
assumed any particular value for $g$.  It is useful to assume
that~(\ref{Sg2}) is written for $g=2$, so that $S_0$ is regular in the
limit $m\to0$.  For a general $g$, we use Eq.~(\ref{SgS2}) to write
\begin{equation}\label{Sg}
  S_g =  \!\int\!d^3x \left(\frac\nu{4\pi}
    \epsilon^{\mu\nu\lambda}a_\mu\d_\nu a_\lambda
  - \sqrt h\, \rho v^\mu D_\mu\varphi 
  + \frac{g-2}{8m} \rho\epsilon^{\mu\nu\lambda} n_\mu
      \tilde F_{\nu\lambda}\right) + S_0[\rho, h^{\mu\nu}, n_\mu, v^\mu].
\end{equation}
Note that the definition of $\tilde A_\mu$ involves $g$, and that the
new term modifies the electromagnetic current:
\begin{equation}\label{jg}
  j^\mu = \frac{\delta S_g}{\delta A_\mu} = \rho v^\mu +
  \frac{g-2}{4m} \varepsilon^{\lambda\mu\nu} n_\lambda \d_\nu\rho .
\end{equation}
The action~(\ref{Sg}) satisfies all conditions outlined in
Sec.~(\ref{sec:reqs}).  From construction, we should regard all the
fields that have been introduced ($\rho$, $v^i$, $a_\mu$, $\varphi$)
as dynamical fields, with respect to which the action is extremized.

Except for one possible topological term, $\int\!\omega d\omega$, all
contributions to $S_0$ depend on microscopic physics, and hence
non-universal.  The term $\int\!\omega d\omega$ does not affect
quantities computed later in Sec.~\ref{sec:physics}.  A full
classification of all possible terms in $S_0$ is beyond the scope of
this paper.  Some of these terms are
\begin{multline}\label{S0form}
  S_0 = -\!\int\!d^3x\sqrt h \bigl[
  \epsilon_i(\rho) + \alpha_1(\rho) h^{\mu\nu} \d_\mu\rho \d_\nu \rho
  + \alpha_2(\rho) h^{\mu\alpha} h^{\nu\beta} 
     \hat\sigma_{\mu\nu} \hat\sigma_{\alpha\beta}
  +\alpha_3(\rho) \sigma^2
   \\
   + \alpha_4(\rho) v^\mu \omega_{\mu\nu} h^{\nu\lambda}\d_\lambda \rho
  + \cdots \bigr],
\end{multline}
where $\epsilon_i$, $\alpha_1$, $\alpha_2$, etc. can be arbitrary
functions of $\rho$.  The function $\epsilon_i$ has the meaning of the
interaction energy of the quantum Hall state with density $\rho$.

Our formalism is applicable equally for gapped FQH states with $\nu<1$
and the IQH state with $\nu=1$.  The difference between them is only
in $S_0$.  For example, by dimensional counting, the coefficients
$\alpha_2$ and $\alpha_3$ in Eq.~(\ref{S0form}) should be proportional
to the the inverse Coulomb gap in the FQH case and the inverse
cyclotron energy in the IQH case.

Putting Eq.~(\ref{Sg}) to flat space we find the Lagrangian
\begin{equation}\label{S-a}
  \mathcal L = \frac\nu{4\pi}
  \epsilon^{\mu\nu\lambda} a_\mu \d_\nu a_\lambda
  + \frac{m\rho v^2}2 
  - \rho D_0\varphi - \rho v^i D_i\varphi
  + \frac{s-1}2\rho\nabla\times v
     + \frac{g-2}{4m}\rho B + S_0[\rho, v^i],
\end{equation}
where in this equation $D_\mu\varphi=\d_\mu\varphi-A_\mu+a_\mu$,
$\nabla\times v\equiv\epsilon^{ij}\d_i v_j$.  The field equations can
be obtained by varying the action,
\begin{subequations}\label{hydro-eq}
\begin{align}
  &\rho = \frac\nu{2\pi} b\,, \label{rho-eq}\\
  &\rho v^i = \frac\nu{2\pi}\epsilon^{ij}e_j \,,\label{j-eq}\\
  &\frac{mv^2}2 - D_0\varphi - v^i D_i\varphi + \frac{s-1}2\nabla\times v
    + \frac{g-2}{4m}B + \frac{\delta S_0}{\delta\rho} =0\,, \label{a0eq}\\
  & m\rho v_i -\rho D_i\varphi + \frac{s-1}2 \epsilon_{ij}\d_j\rho 
     + \frac{\delta S_0}{\delta v_i} = 0\,. \label{eq4}
\end{align}
\end{subequations}

\subsection{Comparison to the bosonic Chern-Simons theory}

Let us compare the action obtained above with that of the standard
bosonic Chern-Simons theory~\cite{Zhang:1988wy}.  The latter is
summarized by the following Lagrangian,
\begin{equation}
  \mathcal L = \frac\nu{4\pi}\varepsilon^{\mu\nu\lambda} a_\mu \d_\nu a_\lambda
    + i \psi^\+ D_0 \psi - \frac1{2m_*} D_i\psi^\+ D_i \psi - V(\psi^\+\psi).
\end{equation}
Changing variables to $\psi=\sqrt\rho e^{i\varphi}$, it becomes
\begin{equation}
  \mathcal L = \frac\nu{4\pi}\varepsilon^{\mu\nu\lambda}a_\mu \d_\nu a_\lambda
  -\rho D_0 \varphi - \frac\rho{2m_*}|\nabla\varphi|^2  
  - \frac{|\nabla\rho|^2}{2m_*} - V(\rho).
\end{equation}
To bring the Lagrangian to the form similar to Eq.~(\ref{S-a}), we introduce an auxiliary field $v_i$,
\begin{equation}\label{LbCSphi}
  \mathcal L = \frac\nu{4\pi}\varepsilon^{\mu\nu\lambda}a_\mu \d_\nu a_\lambda
  -\rho D_0 \varphi - \rho v_i D_i\varphi + \frac{m_*\rho}2 v^2 
  - \frac{|\nabla\rho|^2}{2m_*} - V(\rho).
\end{equation}
One can see some similarities with Eq.~(\ref{S-a}), but there are
obvious differences.  One difference is the term containing $s$ in
Eq.~(\ref{S-a}).  As it is clear from the calculations in the next
Section, this term is essential to reproduce correct next-to-leading
order corrections to electromagnetic response at finite wave numbers.
The field equations following from Eq.~(\ref{LbCSphi}) can be
interpreted as the Euler hydrodynamic equation of a fluid with a
constraint relating the density and the vorticity~\cite{Stone:1990cd}.
Equations~(\ref{hydro-eq}) can also be recast as hydrodynamic
equations, but the form of the equations is more complicated and we
will not write them down here.  We only note that these equations
contain the information about the Hall viscosity, absent in the Euler
equation of Ref.~\cite{Stone:1990cd}.

\section{Physical consequences}
\label{sec:physics}

We discuss some implications of the hydrodynamic theory.  We will
concentrate on linear response to external field.

\subsection{Gravitational response: Hall viscosity}
Let us turn on a weak time-dependent, spatially uniform, and
traceless gravitational perturbation,
\begin{equation}
  h_{ij} = \delta_{ij} + \tilde h_{ij}(t),\quad \tilde h_{ii} = 0.
\end{equation}
Due to rotational symmetry, such a perturbation cannot excite, to
linear order, perturbations of $\rho$ and $v^i$.  The action then
reduces to
\begin{equation}
  S = s\rho\! \int\!d^3x\, \omega_0 = \frac12 s\rho\!\int\!d^3x\,
      \epsilon^{jk}\tilde h_{ij} \d_0 \tilde h_{ik}\,,
\end{equation}
from which we find that the Hall viscosity is
\begin{equation}
  \eta_{\rm H} = \frac{\rho s}2 = \frac{\rho \mathcal{S}}4\,.
\end{equation}
This relationship was derived in Ref.~\cite{ReadRezayi:2010}.

\subsection{Electromagnetic response: preliminaries}
We now assume the space is flat, and discuss linear response to
electromagnetic perturbations.  Such linear response is parameterized by
the polarization tensor $\Pi^{\mu\nu}$,
\begin{equation}
  j^\mu(\omega,\q) = \Pi^{\mu\nu}(\omega,\q) A_\nu(\omega,\q).
\end{equation}
Current conservation restricts the form of $\Pi^{\mu\nu}$ to three
independent functions, $\Pi_{0,1,2}(\omega,\q)$,
\begin{align}
  \Pi^{00} &= q^2 \Pi_0 ,\\
  \Pi^{0i} &= \omega q_i \Pi_0 - i\epsilon^{ij} q_j \Pi_1,\quad
  \Pi^{i0} = \omega q_i \Pi_0 + i\epsilon^{ij} q_j \Pi_1\\
  \Pi^{ij} &= \omega^2\delta^{ij}\Pi_0 + i\epsilon^{ij} \omega\Pi_1
              + (q^2\delta^{ij}-q^iq^j) \Pi_2. 
\end{align}

In order to find $\Pi^{\mu\nu}$ we consider small perturbations
\begin{equation}
   B = B_0 + \tilde B, \quad E_i = 0 + E_i, \qquad
  \rho = \rho_0 + \tilde \rho, \quad v^i = 0 + v^i, \textrm{etc.}\\
\end{equation}
where $\rho_0=\nu B_0/2\pi$.  The linearized equation can be written
as
\begin{align}
  & \tilde\rho + \frac{(s-1)\nu}{4\pi\rho_0}\nabla^2\tilde\rho
  = \frac\nu{2\pi}\left( \delta B + m\nabla\times v + \frac1{\rho_0}
    \vec\nabla \times \frac{\delta S_0}{\delta \vec v}\right),\\
  & m\dot v_i -\epsilon_{ij}B_0 v_j  - \frac{s-1}2 [
    \epsilon_{ij}\d_j(\nabla\cdot v) + \d_i(\nabla\times v)]
    +\frac1{\rho_0}\d_t \frac{\delta S_0}{\delta v_i}
    - \d_i \frac{\delta S_0}{\delta\rho} = E_i + 
   \frac{g-2}{4m}\d_i B . \label{NS}
\end{align}
After solving these equations for $\tilde\rho$ and $v^i$, the current
can be computed from Eq.~(\ref{jg}).  Clearly, a full calculation of
$\Pi^{\mu\nu}$ requires a knowledge of $S_0$.  However, certain
statements about the behavior of $\Pi_{0,1,2}$ at small $q$ can be
made without knowing the coefficients of terms appearing in
$S_0$ [Eq.~(\ref{S0form})].  We will present the results, mostly
without detailed derivations, as they follow in a quite
straightforward manner from the linearized equations above.

\subsection{$\Pi_0$, Kohn's theorem, and static susceptibility}
First, for $\Pi_0$
\begin{equation}
  \Pi_0 = \frac{\nu m}{2\pi} \frac{B_0}{B_0^2-m^2\omega^2} + O(q^2).
\end{equation}
The fact that the $\Pi_0$ is completely determined at $q=0$ is the
content of Kohn's theorem: the response of the system to homogeneous
electric field is independent of interactions.
It is still instructive to
derive Kohn's theorem directly from the field equations.  Consider a situation
when the magnetic field is uniform, and the electric field is uniform
and time-dependent $\mathbf{E}(t)$.  In this case we expect $\rho$ to
remains constant and $v^i=v^i(t)$.  Eq.~(\ref{NS}) now becomes
\begin{equation}
  m\dot v_i = E_i + \epsilon_{ij}  v_j B,
\end{equation}
which is just the equation of motion of the center of mass, which is
independent of interactions.

The static susceptibility is 
\begin{equation}
  \chi(q) = - \Pi^{00}(0,q) = - \frac{\nu m}{2\pi B_0}q^2 + O(q^4).
\end{equation}
In the limit $m\to0$ the $q^0$ part of $\Pi_0$ vanishes, as the $q^2$
term in $\chi(q)$.  The first nonzero contribution ($q^2$ in $\Pi_0$
and $q^4$ in the static susceptibility) comes from the $\hat\sigma^2$
term in $S_0$.

\subsection{$\Pi_1$, density in inhomogeneous magnetic field, and Hall 
conductivity at finite wavenumbers}

Next, for $\Pi_1$ we expand over $q^2$,
\begin{equation}
  \Pi_1(\omega,q) = \Pi_1^{(0)}(\omega) + (q\ell_B)^2\Pi_1^{(2)}(\omega),
  \quad \ell_B = \frac1{\sqrt B} \,.
\end{equation}
The $q^0$ part is determined completely
\begin{equation}
  \Pi_1^{(0)}(\omega) = \frac\nu{2\pi} \frac{B^2}{B^2-m^2\omega^2}\,.
\end{equation}
but $\Pi_1^{(2)}$ is universal only at zero frequency, and narrowly
speaking only in the limit $m\to0$.  For a nonzero $m$ we need to know
the interaction energy as a function of $\rho$, $\epsilon_i(\rho)$,
\begin{equation}
  \Pi_1^{(2)}(0) = \frac\nu{2\pi} \left[ \frac s2 -1 + \frac g4
    - \frac{\nu m}{2\pi}\epsilon_i''(\rho_0) \right].
\end{equation}

There are two physical predictions one can draw from $\Pi_1$.  For
simplicity let us take the limit $m\to0$.  The first prediction is a
formula giving the number density in an static inhomogeneous magnetic
field,
\begin{equation}\label{rho-B}
  \rho = \frac\nu{2\pi}B - \left(\frac s2 -1 + \frac g4\right)\nabla^2
         \ln B + O(\nabla^4).
\end{equation}
One can show that this relationship, as written, remains valid when the
variation of $B$ is not small but of order 1.

The second prediction is for $\sigma_{xy}$ at finite wavenumber $q$.
We assume that the magnetic field $B$ is constant and there is a
static scalar potential $A_0(\x)$, inducing a static longitudinal
electric field $\mathbf E = \bm{\nabla}A_0$.  The Hall current is
\begin{equation}\label{sxy-q2}
  \mathbf{j} = \frac\nu{2\pi} \left[ \mathbf{E} - \left(
   \frac s2 -1 + \frac g4 
  \right) \ell_B^2\nabla^2 \mathbf{E}\right]\times \mathbf{\hat z}.
\end{equation}
In particular, when $g=0$ the result of Ref.~\cite{Hoyos:2011ez} is
reproduced.  Formulas similar to Eqs.~(\ref{rho-B}) and (\ref{sxy-q2})
were obtained in Refs.~\cite{Wiegmann-Kirchhoff,Abanov}.  In fact, the
action~(\ref{S-a}) without the $S_0$ term coincides with the one
proposed in Ref.~\cite{Abanov}.

\subsection{$\Pi_2$ and current in static inhomogeneous magnetic field}

$\Pi_2$ is singular if $g\neq2$ in the limit $m\to0$, and the leading
$m^{-1}$ behaviors of the $q^0$ and $q^2$ terms, at zero frequency,
are universal,
\begin{equation}
  \Pi_2(0,\q) = - \frac{(2-g)\nu}{4\pi m} \left[ 
  1 + \left(\frac s2 - \frac34 + \frac g8\right)(q\ell_B)^2
  \right]  + O(m^0, q^4) .
\end{equation}
This expression determines the current in a static inhomogeneous
magnetic field,
\begin{equation}
 j^i = - \frac{(2-g)\nu}{4\pi m} \epsilon^{ij}
   \left[ 
  1 - \left(\frac s2 - \frac34 + \frac g8\right)\ell_B^2\nabla^2 \right]
  \d_j B .
\end{equation}
For $g=2$, however, the first term in $\Pi_2(0,q)$ is
$m^0$ and not universal.

The results that we derived is valid for any gapped quantum Hall
states with Galilean invariance.  Thus, they can be verified for the
simplest integer quantum Hall state of noninteracting electrons with
$\nu=1$.  For this case, the polarization tensor has been computed
previously~\cite{Chen:1989xs}, and the results can be checked to agree
with our results when one puts in the latter $g=0$ and $s=1/2$.

\section{Conclusion}
\label{sec:conclusion}

In this paper we have shown how one can construct an effective field
theory of the quantum Hall state which respects nonrelativistic
diffeomorphism invariance.  The most convenient mathematical framework
turns out to be the Newton-Cartan geometry, previously considered in
the literature in a different context.  One of the most attractive
features of the formalism is regularity of the massless limit $m\to0$.
The action cleanly separates universal physics from non-universal
physics. The latter is is parameterized by an action $S_0$, whose form
is restricted by the Newton-Cartan symmetry.  Even without any
dynamical information, one can already make several predictions,
including the $q^2$ correction to the static Hall conductivity.

In this paper we have concentrated our attention to the regime of very
low frequencies.  To treat the physics at the scale of the Coulomb
gap, we need to know more information about the non-universal part of
the action, $S_0$.  Knowing that $S_0$ depends on the Coulomb energy
allows us to conclude, for example, that the $q^2$ part of the Hall
conductivity $\Pi_1$ has nontrivial frequency dependence at the
Coulomb energy scale.  This is consistent with the results of
Ref.~\cite{Bradlyn:2012ea}.

In our formalism, among the components that make up the Newton-Cartan
geometry, only the velocity $v^i$ is treated as a dynamical variable.
The metric $h_{ij}$ is simply the background metric.  On the other
hand, it has been suggested recently~\cite{Haldane:2011ia} that some
``internal metric'' plays the role of a dynamic degree of freedom in
FQH systems.  We hope that future investigations will elucidate the
connection between our approach and the suggestion of
Ref.~\cite{Haldane:2011ia}.

Finally, the implications for edge modes, for quasiholes and
quasiparticles need to be investigated.  The insights that we obtain
in this paper may be useful for the construction of holographic models
of quantum Hall systems.

\acknowledgments

The author thanks A.~Abanov, I.~Gruzberg, M.~Levin, E.~Martinec,
S.~Simon, and P.~Wiegmann for discussions.  This work is supported, in
part, by DOE grant DE-FG02-90ER-40560 and NSF MRSEC grant DMR-0820054.

\end{document}